\documentstyle[12pt,epsfig,amsmath,amssymb]{article}
\begin{document}

\title{\bf Quantum Darwinism requires an extra-theoretical assumption of encoding redundancy}

\author{{Chris Fields}\\ \\
{\it 21 Rue des Lavandi\`eres}\\
{\it Caunes Minervois 11160 France}\\ \\
{fieldsres@gmail.com}}
\maketitle

\begin{abstract}
Observers restricted to the observation of pointer states of apparatus cannot conclusively demonstrate that the pointer of an apparatus $\mathcal{A}$ registers the state of a system of interest $S$ without perturbing $S$.  Observers cannot, therefore, conclusively demonstrate that the states of a system $S$ are redundantly encoded by pointer states of multiple independent apparatus without destroying the redundancy of encoding.  The redundancy of encoding required by quantum Darwinism must, therefore, be assumed from outside the quantum-mechanical formalism and without the possibility of experimental demonstration.

\end{abstract}

\textbf{Keywords} Quantum Darwinism, Environment as witness, Redundant encoding

\section{Introduction}

The concept of ``quantum Darwinism" was introduced by W. Zurek to emphasize, by analogy with the criterion of reproductive efficiency imposed by biological Darwinism, the central role in the emergence of classicality played by the redundant encoding of einselected pointer states of a quantum system $S$ by orthogonal (for all practical purposes) fragments of the environment $\mathcal{E}$ \cite{zurek03rev, zurek09rev}.  In a recent and comprehensive review \cite{zurek09rev}, Zurek shows that redundant encoding of pointer states by $\mathcal{E}$ not only enables but forces multiple observers who agree about which observables to measure, and who conduct independent measurements on orthogonal fragments of $\mathcal{E}$, to agree on the measured values of the chosen observables.  The consistency forced on observers by $\mathcal{E}$'s role as a witness to the evolution of $S$ enables both an explication of the appearance of wavefunction collapse and a derivation of Born's rule using what is claimed to be a purely quantum-mechanical concept of probability \cite{zurek09rev}.  These results have been extended to cases in which the environment is noisy, and hence to cases of measurements with experimental uncertainty \cite{zwolak09}.

This brief note does not challenge any of the formal results obtained in \cite{zurek09rev}.  However, it does point out that the concept of redundancy of encoding upon which quantum Darwinism rests is implicitly and irreducibly classical.  It shows that multiple observers making independent measurements cannot demonstrate that their measurements access redundant encodings of information about a single system $S$, but can at best assume redundancy of encoding on the basis of classical observations alone.  Hence quantum Darwinism is circular: it requires an assumption of encoding redundancy justified by classical observations to explain the emergence of classicality.  Although quantum Darwinism significantly illuminates the process by which classicality emerges, it therefore does not provide an explication of the emergence of classicality in purely quantum-mechanical terms, and hence does not achieve the conceptual advance over Bohr's insistance that apparatus be viewed as intrinsically classical that it appears, at first glance, to represent.

\section{Encoding by entanglement is redundant but ambiguous}

Zurek defines the redundancy $R^{S}$ of information about a given system $S$ available in $\mathcal{E}$ as $R^{S}_{\delta} = \frac{1}{f_{\delta}}$, where $f_{\delta}$ is the size of a fragment $\mathcal{F}$ of the environment that encodes all but $\delta$ of the information about the pointer states of $S$ (Eq. 4 of \cite{zurek09rev}).  Critical to this definition is the stipulation that the fragments $\mathcal{F_{\mathit{i}}}$ under consideration, each of which is taken to have a size on the order of $f_{\delta}$, be disjoint and not coherently entangled, and hence dynamically independent for all practical purposes.  It is the dynamical independence of the fragments $\mathcal{F_{\mathit{i}}}$ that renders the encoding of pointer-state information redundant, as it permits one or more observers to independently gather information about the pointer states of $S$ without perturbing either $S$ or each other's records.  Observers of different fragments interact with distinct, disjoint and independent encodings of the pointer states of $S$, e.g. distinct and independent substates of the photon field.  Because such distinct environmental substates encode the same pointer state of $S$ at any given time, observers interacting with such substates can and must agree not only about $S$ being in some pointer state (i.e. they agree that $|S>$ has ``collapsed") but also about which pointer state $S$ is in (i.e. they can collaboratively derive the Born Rule).  Hence redundant encoding appears to ``largely settle" the measurement problem (\cite{zurek09rev} p. 187).

The above definition of redundancy requires that observers can distinguish the encodings, in their fragments of the environment, of the pointer states of distinct systems.  It requires, in other words, that if $S$ and $S^{\prime}$ are distinct systems embedded in $\mathcal{E}$, observers will be able to choose fragments $\mathcal{F_{\mathit{i}}}$ of $\mathcal{E}$ in such a way that the pointer states $\{|j>\}$ and $\{|j^{\prime}>\}$ of $S$ and $S^{\prime}$ respectively will be distinguishable in each of the $\mathcal{F_{\mathit{i}}}$.  If but only if this assumption is true, observers detecting indistinguishable pointer states encoded by their distinct fragments of $\mathcal{E}$ will be justified in concluding that their pointer states are redundant encodings that indicate the existence of a unique observed system.  This assumption that distinct systems will encode distinguishable pointer states is an instance of ``Leibnitz's law" of the identity of indiscernibles \cite{forrest06}, where the indiscernibles in question are the independently-observed environmental encodings of the pointer states of each distinct system.  In order for quantum Darwinism to explain the emergence of classicality in quantum-mechanical terms, this assumption must not only be true, it must be demonstrably true from within quantum mechanics.

The difficulty with quantum Darwinism is that redundancy of encoding cannot, in fact, be demonstrated.  Consider a situation in which macroscopic observers $O_{1}$ and $O_{2}$ have access to two disjoint macroscopic fragments $\mathcal{F_{\mathrm{1}}}$ and $\mathcal{F_{\mathrm{2}}}$ of the environment $\mathcal{E}$.  Each fragment $\mathcal{F_{\mathit{i}}}$ contains a macroscopic apparatus $\mathcal{A_{\mathit{i}}}$, which following Tegmark \cite{tegmark} is regarded as consisting solely of a pointer $\mathcal{P_{\mathit{i}}}$ that can indicate any of $N$ pointer states $|\mathcal{P_{\mathit{i}}^{\mathit{j}}}>$.  Suppose the observers each independently observe their pointers, and record indistinguishable values; for example, each pointer indicates the value ``5".  The observers then confer.  Are $O_{1}$ and $O_{2}$ justified in concluding that $\mathcal{P_{\mathrm{1}}}$ and $\mathcal{P_{\mathrm{2}}}$ are redundantly registering the information that a single system $S$ is in a pointer state $|5>$ that its interaction with $\mathcal{E}$ has redundantly entangled with both $|\mathcal{P_{\mathrm{1}}^{\mathrm{5}}}>$ and $|\mathcal{P_{\mathrm{2}}^{\mathrm{5}}}>$?  That is, can they rule out on the basis of observations the alternative possibility that two distinct, dynamically-decoupled systems $S$ and $S^{\prime}$ are interacting with the environment, and that both are independently in pointer states $|5>$ and $|5^{\prime}>$ that the available apparatus indicate by registering, each independently with its own pointer, the value ``5"?  Figure 1 illustrates this dilemma.
  
\begin{figure}[hbt]
\epsfig{file=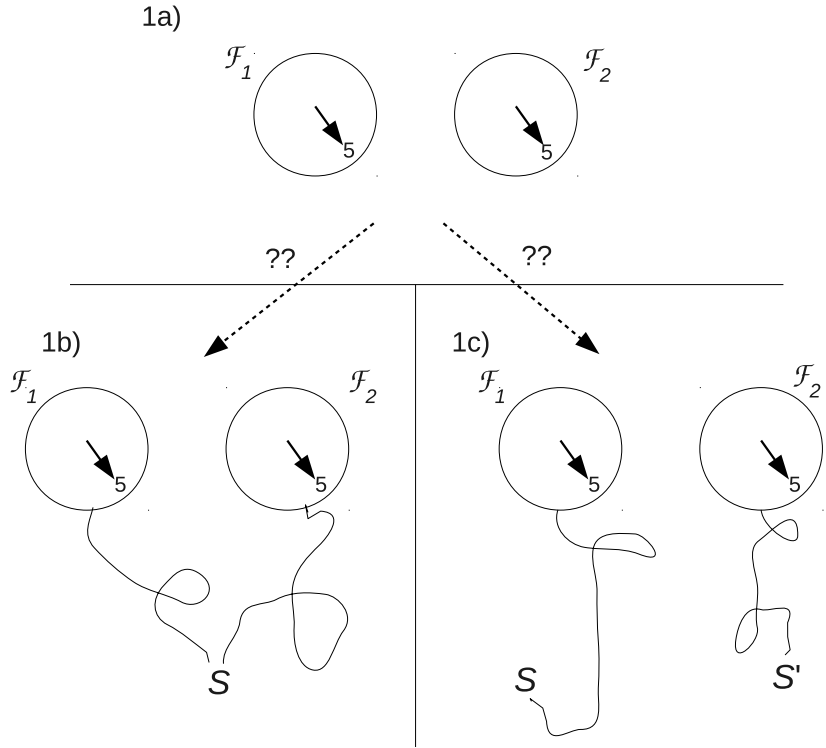, height=4.5in}
\caption{
Indistinguishable pointer states, e.g. pointer values of ``5" independently registered by apparatus embedded in distinct fragments $\mathcal{F_{\mathrm{1}}}$ and $\mathcal{F_{\mathrm{2}}}$ of the environment (Panel 1a) may be redundant encodings of the state of a single system $\mathit{S}$ (Panel 1b) or coincidental encodings of the states of two distinct systems $\mathit{S}$ and $\mathit{S^{\prime}}$ (Panel 1c).  Demonstrating redundancy by ruling out the situation shown in Panel 1c requires interacting directly with $\mathit{S}$, hence destroying redundancy.   
}
\end{figure}

The dilemma illustrated in Fig. 1 is not far-fetched.  One need only imagine that the apparatus in $\mathcal{F_{\mathrm{2}}}$ is being calibrated, without either observer's knowledge, by an unseen technician who happens to choose ``5" as a calibration value.  More seriously, the apparatus in $\mathcal{F_{\mathrm{1}}}$ and $\mathcal{F_{\mathrm{2}}}$ may register apparently confirmatory values that prove, on further investigation, to be entirely spurious.  Such situations are not unknown in the history of particle physics; the claims and counterclaims regarding the pentaquark (reviewed in \cite{hicks}) provide a recent example.  The relevant question is \textit{how} one goes about demonstrating redundant encoding or the lack thereof.  The answer is suggested in Fig. 1: one traces the wires until the physical interactions generating the pointer states $|\mathcal{P_{\mathrm{1}}}>$ and $|\mathcal{P_{\mathrm{2}}}>$ are fully revealed.  Only if these interactions can be shown to couple the independent pointers of $\mathcal{A_{\mathrm{1}}}$ and $\mathcal{A_{\mathrm{2}}}$ to the same system $S$, i.e. to a single subset of quantum degrees of freedom, can the encoding be shown to be redundant.

``Tracing the wires" involves expanding what counts as ``apparatus" at the expense of what counts as ``environment" within each fragment $\mathcal{F_{\mathit{i}}}$.  Each observer $O_{i}$ interacts with a sequence of progressively higher-dimensional pointer states $|\mathcal{P_{\mathit{i}}^{\mathrm{5}}}, \mathit{x_{i}^{1}}>$, $|\mathcal{P_{\mathit{i}}^{\mathrm{5}}}, \mathit{x_{i}^{1}, x_{i}^{2}}>$, ..., $|\mathcal{P_{\mathit{i}}^{\mathrm{5}}}, \mathit{x_{i}^{1}, x_{i}^{2}, ..., x_{i}^{n}}>$ that include progressively more positions $x^{j}$ of the wire; in actual practice, these expanded pointer states incorporate readings from larger and larger subsets of available apparatus.  Any such sequence of progressively-expanded pointer states terminates at a quantum limit beyond which the incorporation of additional environmental degrees of freedom into the observed apparatus irreversibly perturbs the system.  At this quantum limit, decoherence fails; the quantum degrees of freedom ``inside the box" are entangled with the apparatus components with which they interact, and are distinguishable from such apparatus components only by notational convention.  Determining which apparatus degrees of freedom are entangled at amplitudes sufficient for observation with which system degrees of freedom at this quantum scale is not possible without destroying the entangled quantum state.  Hence $O_{1}$ and $O_{2}$ cannot determine whether their respective pointer states are generated by interactions between their respective apparatus $\mathcal{A_{\mathrm{1}}}$ and $\mathcal{A_{\mathrm{2}}}$ and a single subset of quantum degrees of freedom.  They cannot, in other words, demonstrate redundancy of encoding.  

In practice, $O_{1}$ and $O_{2}$ may agree that their expanded apparatus pointer states $|\mathcal{P_{\mathit{i}}^{\mathrm{5}}}, \mathit{x_{i}^{1}, x_{i}^{2}, ..., x_{i}^{n}}>$ provide sufficient evidence for them to \textit{assume} that their original pointer states $|\mathcal{P_{\mathit{i}}^{\mathrm{5}}}>$ were generated by redundantly encoded information indicating a unique system state $|5>$.  Such an assumption is, however, based on classical observations and introduced from outside the quantum framework.  Hence while Zurek's analysis shows that for a suitable choice of $\mathcal{F_{\mathit{i}}}$, redundant registration of the pointer states of a pre-selected system $S$ by the independent pointers of multiple apparatus $\mathcal{A_{\mathit{i}}}$ embedded in the $\mathcal{F_{\mathit{i}}}$ is \textit{possible}, it does not show that redundant registration of $|S>$ by any
combination of the apparatus pointers is \textit{necessary}.  Nothing in the quantum-mechanical formalism prevents the pointer of a given apparatus $\mathcal{A_{\mathit{k}}}$ embedded in a fragment of the environment $\mathcal{F_{\mathit{k}}}$ from registering the state of some alternative, decoupled system $S^{\prime}$ that interacts with $\mathcal{E}$ so as to generate a pointer state $|\mathcal{A_{\mathit{k}}}>$ indistinguishable from those of the other $\mathcal{A_{\mathit{i}}}$, regardless of what information about $S$ may be present in $\mathcal{F_{\mathit{k}}}$.  A party of observers can agree to treat only $S$ as a system of interest, and can assume that the available apparatus redundantly registers its states, but nothing in the quantum mechanical formalism or in the data available through observation enforces such agreements or assumptions.  Quantum mechanics does not dictate system-environment decompositions, and does not enforce Occam's razor or Leibnitz's law.  Indeed the principle of superposition explicitly forbids such notional pre-selections of systems of interest from having physical consequences at scales where entanglement is relevant.  The assumption that a tree viewed from two different angles is the same system of interest is not forced on one by quantum mechanics.  It is a classical assumption motivated by our bias toward object permanence, and by practical experience at macroscopic scales.

\section{Conclusion}

Observers have access only to the pointer states of apparatus.  Indistinguishable apparatus pointer states suggest, but do not demonstrate, redundant encoding.  While theoretical and practical considerations may provide excellent evidence to justify an expectation that a given fragment $\mathcal{F_{\mathit{i}}}$ of the environment encodes information about a pre-selected system of interest $S$, and that an apparatus $\mathcal{A_{\mathit{i}}}$ embedded in $\mathcal{F_{\mathit{i}}}$ registers that information with its pointer, this situation cannot be demonstrated without destroying redundancy.  Redundant encoding of system pointer states by apparatus pointer states can at best be assumed if redundency is to be maintained.  Such an assumption is motivated and justified by classical, not quantum-mechanical considerations.

Quantum Darwinism cannot, therefore, be regarded as fully accounting for the appearance of a macroscopic world of redundantly encoded and hence re-identifiable objects.  It can be claimed with confidence that ``quantum states acquire objective existence when reproduced in many copies" (\cite{zurek09rev} p. 165); the formalism of quantum Darwinism demonstrates this.  However, the ``objective existence" rests on the assumption that the ``many copies" are in fact copies.  This is a classical assumption for which conclusive experimental evidence cannot be obtained.

\section*{Acknowledgement}

The suggestions of an anonymous referee were helpful in improving the clarity of this paper.

\end{document}